\documentclass{iaus}
\usepackage{graphicx}
\title[From Interacting Binaries to Exoplanets: Essential Modeling Tools] 
{New Galactic Candidate Luminous Blue \\ Variables and Wolf-Rayet Stars}

\author[Guy S. Stringfellow, Vasilii V. Gvaramadze, Yuri Beletsky, \and Alexei Y. Kniazev]   
{Guy S. Stringfellow$^1$, Vasilii V. Gvaramadze$^2$, Yuri Beletsky$^3$, 
 \and Alexei Y. Kniazev$^4$}

\affiliation{$^1$Center for Astrophysics and Space Astronomy, \\
University of Colorado, 389 UCB, Boulder, CO 80309-0389, USA 
\\ email: {\tt Guy.Stringfellow@colorado.edu} \\[\affilskip]
$^2$Sternberg Astronomical Institute, Moscow State University, Universitetskij Pr. 13, Moscow 119992, Russia 
\\email: {\tt vgvaram@iki.rssi.ru} \\[\affilskip]
$^3$European Southern Observatory, Alonso de Cordova 3107, Santiago, Chile \\email: {\tt ybialets@eso.org} \\[\affilskip]
$^4$South African Astronomical Observatory, PO Box 9, 7935 Observatory, Cape Town, South Africa
\\email: {\tt akniazev@saao.ac.za} 
}

\pubyear{2012}
\volume{282}  
\pagerange{XXX}
\jname{From Interacting Binaries to Exoplanets: Essential Modeling Tools}
\editors{Mercedes Richards \& Ivan Huben\'{y}, eds.}
\begin{document}

\maketitle

\begin{abstract}
We have undertaken a near-infrared spectral survey of stars
associated with compact mid-IR shells recently revealed 
by the MIPSGAL (24 $\rm \mu$m) and GLIMPSE (8 $\rm \mu$m)
{\it Spitzer} surveys, whose morphologies
are typical of circumstellar shells produced by massive evolved 
stars. Through spectral similarity with known Luminous Blue Variable (LBV) 
and Wolf-Rayet (WR) stars, 
a large population of candidate LBVs (cLBVs) and a smaller 
number of new WR stars are being discovered. This significantly increases the 
Galactic cLBV population and confirms that 
nebulae are inherent to most (if not all) objects of this class. 
\keywords{stars: emission-line, Be, stars: mass loss, stars: winds, outflows, stars: Wolf-Rayet}
\end{abstract}

\section{Introduction}
Despite intensive search efforts over the last several decades the Galactic Luminous Blue Variable (LBV) 
population has remained sparse. This paucity is difficult to reconcile with our understanding of stellar evolution of the most massive stars. Until the last year or two there were only 12 confirmed Galactic 
LBVs known, and 23 candidate-LBVs (\cite[Clark et al. 2005]{Clark2005}). LBVs display 
rather unique rich infrared emission line spectra, including contributions from H, He, Mg\,{\sc ii}, Na\,{\sc i}, and
Fe\,{\sc ii}.
%and often [Fe\,{\sc ii}]. 
Visual inspection of the {\it Spitzer} GLIMPSE (\cite[Benjamin et al. 2003]{Benjamin2003}) and MIPSGAL
(\cite[Carey et al. 2009]{Carey2009}) Galactic plane surveys have produced catalogues of previously unknown 8 $\rm \mu$m and 24 $\rm \mu$m
nebulae with concentric point sources that can be traced back to 2MASS $K$-band or even optical sources as possible progenitors of the associated nebulae (\cite[Gvaramadze et al. 2010a]{Gvar2010a}, \cite[Wachter et al. 2010]{Wachter2010}), especially with new imaging (\cite[Stringfellow et al. 2012]{String2012}). We are conducting a near-IR spectral survey to identify 
new cLBVs and WRs that produced these shells.
\begin{figure}[t]
\begin{center}
 \includegraphics[width=5.0in]{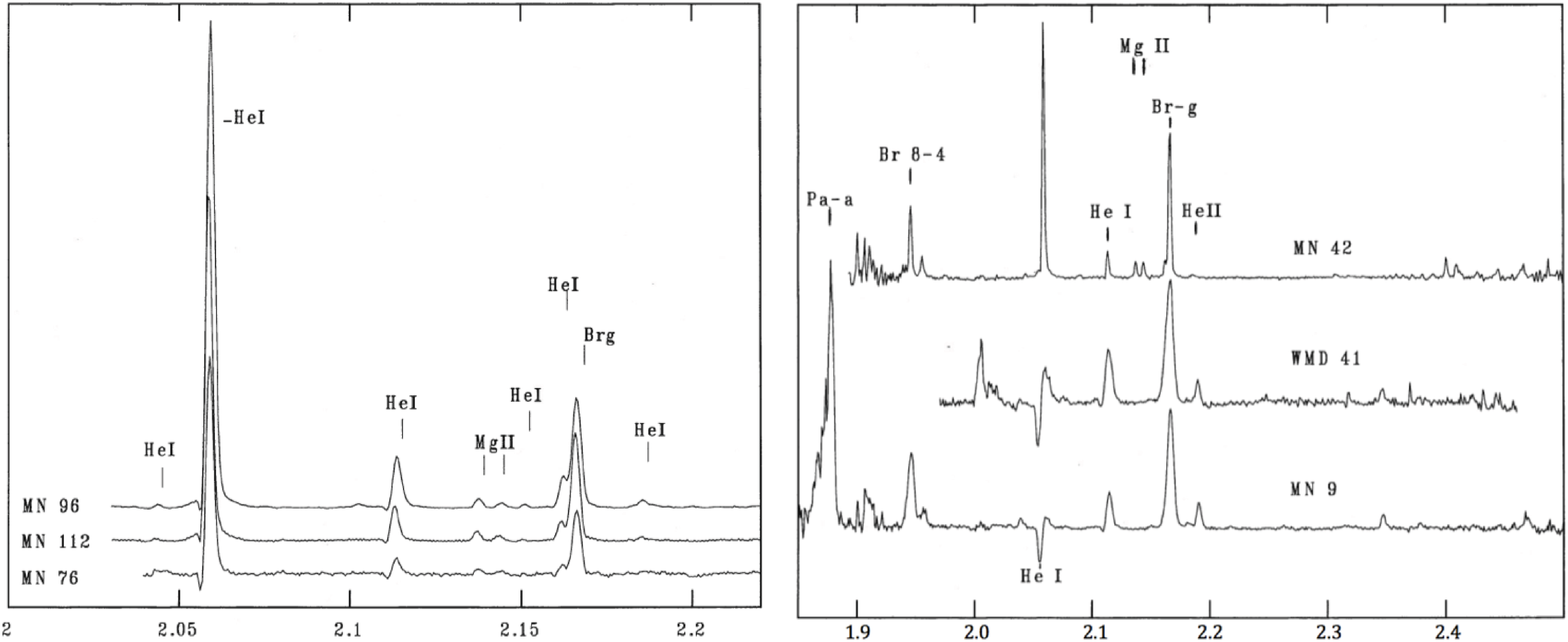} 
 \caption{Normalized $K$-band spectra of newly identified cLBVs and WRs. See Figure 1 of
\cite[Stringfellow et al. 2012]{String2012} for images identifying MN 96, including optical 
recovery in the $I$-band.}
   \label{fig1}
\end{center}
\end{figure}

\section{Observations and Results}
We have obtained spectra of $\sim$50 stars associated with newly discovered mid-IR nebulae using SpeX 
on the NASA IRTF 3m, Triplespec on APO 3.5m and Palomar Hale 5m, and ISAAC on the ESO-VLT. A few of 
the $K$-band spectra are shown in Figure 1. The left panel shows three newly identified cLBVs that have Fe\,{\sc ii} emission 
absent in their IR spectra. 
%that do {\it not} display Fe II in their IR spectra. 
Prominent line emission arise from He\,{\sc i}, Br\,$\gamma$, and Mg\,{\sc ii}. A spectrum of MN 96 (WMD 54, \cite[Wachter et al. 2010]{Wachter2010}), is discussed in 
\cite[Wachter et al. (2011)]{Wachter2011}, who notes the similarity between both LBV and WR 
late-type WN spectra for this particular star. Our spectrum clearly indicates the absence 
of any He\,{\sc ii} 2.189 $\rm \mu$m line emission. 
Comparison of an optical spectrum of MN 112 with that 
of P Cyg rendered classification as a cLBV
(\cite[Gvaramadze et al. 2010b]{Gvar2010b}); both spectra display numerous 
optical Fe\,{\sc iii} lines, but no Fe\,{\sc ii} line emission. 
The absence of the 2.089 $\rm \mu$m Fe\,{\sc ii} line in the MN 112 $K$-band 
spectrum is consistent with a higher temperature in this line emitting region.  MN 76 (WMD 38) was classified as a Be star (\cite[Wachter et al. 2011]{Wachter2011}), though no spectrum was shown. Clearly the $K$-band spectra for these three stars - MN 96, MN 112, and MN 76 - are nearly identical (barring small differences in line widths and strengths), and should render the same IR spectral classification. These stars could be transitional between the LBV and late-WN stars, or have spectral types varying between minimum contraction to maximum expansion, corresponding to hot and cool temperature phases, respectively. 
VLT spectra for two WRs, WMD 41 (WN8-9h) and MN 9 (WN7-9h), are displayed 
in the right panel of Figure 1 along with the VLT spectrum 
for the cLBV MN 42 (WMD 15). The WR spectra lack Mg II emission and display 
broader H and He lines than
the cLBVs. MN 42 was classified as B[e]/LBV by Wachter et al. (2011) though no spectrum was shown. 
MN 42 resembles those cLBVs shown in the left panel, lacking Fe\,{\sc ii} 
emission, strengthening the case that the Fe{\sc ii}-deficient cLBVs may be transitioning to 
late-WN stars. 
We designate MN 42, MN 76, MN 96, and MN 112 as currently Fe{\sc ii}-deficient cLBVs.

GSS thanks the AAS for receipt of an ITG and SmRG, and the IAU for support.

\end{document}